# NewsPad: Designing for Collaborative Storytelling in Neighborhoods

**J. Nathan Matias**
MIT Media Lab
75 Amherst St
Cambridge, MA 02139
jnmatias@mit.edu

**Andrés Monroy Hernández**
Microsoft Research
One Microsoft Way
Redmond, WA 98052
amh@microsoft.com

## Abstract
This paper introduces design explorations in neighborhood collaborative storytelling. We focus on blogs and citizen journalism, which have been celebrated as a means to meet the reporting needs of small local communities. These bloggers have limited capacity and social media feeds seldom have the context or readability of news stories. We present NewsPad, a content editor that helps communities create structured stories, collaborate in real time, recruit contributors, and syndicate the editing process. We evaluate NewsPad in four pilot deployments and find that the design elicits collaborative story creation.

## ACM Classification Keywords
A5.3 Information interfaces and presentation: Group and Organization Interfaces – collaborative computing

## Introduction
News media have historically reported on events in neighborhoods and small towns. As the US news industry declines, local coverage has weakened [10][13]. Social network sites (SNS), such as Twitter, Facebook, and YouTube, offer opportunities to fill this gap by allowing anyone to act as a journalist [5]. Citizen reports tend to be status updates, images, and videos spread across disparate SNS; they are often published as business reviews [2] and time-based feeds [4] rather than structured stories that fulfill the role of journalism to inform communities.

When neighborhoods do have community curators (e.g., neighborhood bloggers), they work largely following the model of traditional news media. Although rarely a full-time job, the work of a single blogger involves composing labor-intensive posts, even if other community members might also be sharing media on SNS. Many neighborhoods do not even have a single blogger.

In this paper, we present a survey of existing literature and commercial technologies. We introduce a system called NewsPad to address needs in community reporting. We informally evaluate NewsPad in four real-life pilot deployments in neighborhoods and other small communities.

## Background

For decades, researchers have developed systems to address the *information needs of neighborhoods*. As early as 1974, Community Memory served as a digital bulletin board in Berkley [11]. Similarly, in 1994 the Blacksburg Electronic Village was designed for residents of a small town in Virginia, with a system that let people edit documents collaboratively [3]. In 2004, i-Neighbors was developed to support discussions about community issues [6]. Recent systems like Whooly create automated feeds of local information [8].

Articles often include contributions from multiple sources, traditionally collected by a journalist. SNS curation systems like Storify retain this single-curator model [4]. Peer-production communities have developed collaborative sourcing practices. For example, Wikipedians often add markings like "[citation needed]" to invite others to contribute references. Newspapers have also experimented with *social recruitment* techniques by sending geolocated requests for photos to freelance journalists [14].

Research on collaborative writing emphasizes the value of using content structure to simplify the act of contribution. For example, Wikipedia's article *structure* makes it easier for people to read and contribute to than less structured alternatives [7].

## NewsPad: Collaborative Storytelling System

Motivated by observations of existing systems, we developed NewsPad, a content editor for cooperative community storytelling that enables anyone to create a story by helping him or her to write a compelling headline, provide a basic story structure, and share it with their community (see Figure 1).

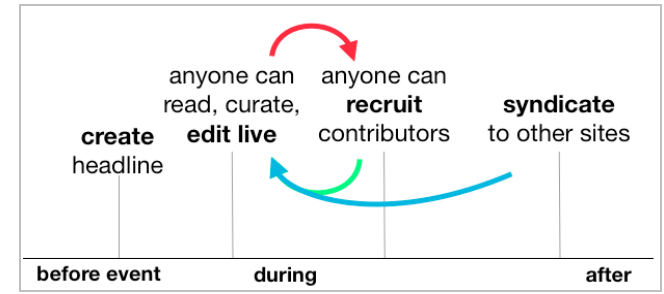

**Figure 1.** The lifespan of a NewsPad story

NewsPad allows multiple readers to edit sections in real time, adding commentary and curating media. People can also leave improvement requests on the story and recruit others to help in specific ways. Stories can be embedded in to third party sites and updated in real time. NewsPad stories can be shared on websites that already function as communication hubs for a community, broadening the pool of readers and potential contributors.

The NewsPad server, written in NodeJS, uses ShareJS to serve stories and synchronize edits across clients. The client application uses Backbone.js and runs on desktop and mobile browsers.

*Establishing Purpose with Structured Stories*
Stories have a headline and a series of sections, each with a heading, embedded media, and accompanying text. Users are guided to create compelling story structures that easily accept contributions from readers. After creators name the event or topic, NewsPad automatically generates compelling headline suggestions. Headlines are based on industry studies that show what titles are most likely to attract attention [12]. A "Zombie Walk" event received the headline suggestions "5 things You Missed at the Zombie Walk" and "Everything You Need to Know About the Zombie Walk." Creators can edit a suggested headline or write their own.

Within each story, headings give each section a purpose that multiple contributors can understand and improve upon. The embedded media area offers space for content curation of photos, video, and microblog content. The text element of a section offers space for the context, commentary, and narrative that make the article a story rather than a feed. All stories have an attribution section where contributors can accept public acknowledgement for their involvement.

*Sharing Effort with Real-time Collaborative Editing*
NewsPad supports real-time collaborative editing within stories. Any reader can make changes that are propagated to all other readers and editors: editing text, adding sections, or reverting to a previous section from the editing history.

To curate content from elsewhere, users click an "embed media" button and insert the URL of a page containing a video, photo, audio, or microblog content. Some versions embed a Twitter stream to support content curation.

*Improving Stories with Social Recruitment*
Readers of NewsPad stories can make *offers to help* and add *improvement requests* to specific parts of a story. Users can add requests to any section of the story, entering their name, a recipient, a specific request, and the topic. People who offered to help are listed as suggested recipients. Requests can also be sent to specific people on Twitter. All requests have a unique link and are added to the story for readers to respond or forward to others.

When recipients click the link, they are asked to make an improvement. Those who accept the request are directed to the specific section that needs improvement. Requests for improvement are dismissed upon completion. A count of outstanding requests is shown to every reader as an invitation to help.

*Reaching Communities with Real-time Syndication*
When NewsPad stories are embedded into third-party websites, they retain full functionality for real-time updates, editing, and social recruitment. Real-time syndication has been tested with Drupal, WordPress, and Tumblr.

**Pilot Deployments**
We evaluated these design ideas for storytelling in neighborhoods with four pilot deployments, starting with a Wizard-of-Oz prototype. Lessons from each test were incorporated into the NewsPad system.

*Social Recruitment at the Capitol Hill Garage Sale*
To test the viability of social recruitment before designing NewsPad, we attended a garage sale in Seattle neighborhood. Three researchers at the event gave people cards requesting email contributions. One researcher searched for the garage sale on Twitter and asked users to submit media. Contributions were added to a garage sale story website.

No one sent a response to the cards. When we asked neighborhood Twitter accounts to submit content, all of them submitted material. Our story with eight sections combined submitted content with photos curated from social media. When we tweeted the post, a community

**Figure 2.** NewsPad Edit View (left), and improvement request (right)

blogger took a screenshot of it, embedded it into his post, and linked to the new post.

This test gave us confidence that electronic social recruitment was a fruitful approach and that curation could be combined with recruitment. The syndication of our content into a community blog inspired our focus on syndication.

*Structured Stories and Real-time Collaborative Editing at a Journalism Conference*
We tested the first NewsPad prototype at a conference talk with thirteen speakers in 90 minutes. This prototype supported real-time collaboration, structured stories, and media curation. Each talk was given a section in the NewsPad story in advance. We posited that, given some initial structure, a good story could be created by multiple people.

Contributors were recruited by (a) asking audience members in person, (b) making an appeal on Twitter, and (c) asking specific Twitter accounts to help. All three strategies yielded contributions. The organizers of the event shared our recruitment tweet, and we observed simultaneous edits in multiple parts of the document. At least five contributors added slide decks, photos, and screenshots of software, as well as notes from each talk. The story was later syndicated to the event hosting organization's blog.

During this test, we discovered needs for attribution, history, a read-only view, and section-specific requests. Attribution shows who to thank and ask to help. History allows for the recovery of previous versions of the stories. A read-only view protects the stories from being accidentally deleted. Section-specific requests allow respondents to find where to contribute.

*Content Curation at a Zombie Walk*
We tested curation from social media at a community Zombie Walk. Participants posted photos and videos to a hashtag on Twitter, a stream that we embedded into the NewsPad editor.

The resulting NewsPad story was mostly comprised of microblog content curated by a researcher during the event. Unlike the conference, where attendees were willing to contribute to a shared story, Zombie Walk participants did not stop during the event to add media to the NewsPad story. By the time people received our tweets, they had moved on. Most photos and videos appeared online several days after the event, and the story grew over days rather than minutes, with potential media continuously aggregated to the editor's hashtag stream.

*Real-time Syndication at a Design Expo*
To test real-time syndication in NewsPad, we reported on short presentations at a design expo. The story was embedded into a Drupal and a Tumblr blog with relevant audiences. Changes instantly propagated to these sites. Readers could click a button to edit the article. After the expo ended, we exported a static version of the story to the two blogs.

Real-time syndication brought several new participants into the editing process and enabled new pathways for the story to be heard. The nine-section story was noticed by the editor of a US-wide broadcaster that sometimes syndicates content from one of the blogs that embedded NewsPad. The broadcaster later published a shortened version of the story.

These pilots highlighted the diversity of approaches for syndication at different times in the life of a story. By embedding the NewsPad editor, we reached a broad pool of contributors. The Tumblr post was shared widely on Facebook. The national broadcaster published an edited, static story, enabling a wider, if smaller audience to encounter it. The live version attracted 135 views on an organization's site and 1052 views on Tumblr. The broadcaster's article received 684 views.

**Future Work**

Future research should evaluate the effectiveness of NewsPad and the patterns it embodies in a broader set of neighborhoods and communities. Pilots revealed different preferences for content production activity across desktop and mobile systems; contributors especially wanted to submit content via email. Future versions should offer more ways for people to contribute to collaborative stories.